# Singular-Turbulent Structure Formation in the Universe and the Essence of Dark Matter II. Large-scale structure formation scenario

Reza Dastvan

*Institute of Biochemistry and Biophysics, University of Tehran, P.O. Box 13145-1384, Tehran, Iran.* dastvan@khayam.ut.ac.ir

ABSTRACT
Based on the proposed unifying theory of dark matter and quintessence, a novel nonlinear structure formation scenario is suggested. This top-down singular and turbulent scenario results in a bottom-up hierarchical clustering and is consistent with various challenging cosmological observations like the existence of massive galaxies and galaxy clusters at very large redshifts ($z \geq 5$). Strong non-linearity is formed in the very early stage, therefore no extra biasing process is needed.

*"...Cosmic shapes are the ocean's foam..."* Mawlānā Jalal al-Din Balkhi (1207-1273 AD) in *Divan-e-Shams* [mystical poems]

## 1 Introduction

In the mean time, despite rapid advancement in theoretical cosmology development, there are certain issues that remain unexplainable in the presently available theories; one of these issues concern the origin and nature of gravitational instability (Coles 2002; Gibson 1999). Recent studies that have incorporated condensation and void formation occurring on the non-acoustic density nuclei produced by turbulent mixing appear to indicate that the universe is inherently *nonlinear* nature.

1.1 Observations to be explained

The epoch $z_d \approx 1000$ is the order of magnitude of the epoch of decoupling. Since the radiation was in equilibrium with the plasma at $T_d$, the uniformity of the cosmic microwave background (CMB) radiation implies that the gas of Hydrogen was also almost perfectly uniform at that time. In contrast, the observed structure of luminous matter is strongly clustered, intermittent, and fractal-like, with correlations over perhaps hundreds of millions of light years.

In the last two decades there have been a number of observations affecting galaxy formation and large-scale structure that have been a potential problem for traditional models which invoked early random Gaussian fluctuations. In particular, many of the advocates of gaussian fluctuations and cold dark matter have tried to argue that these observations are statistical flukes that have yet to be established. Obviously, if these potential observations continue to hold up and are verified and are shown to be ubiquitous rather than statistical rarities, then the traditional models are in serious trouble. Perhaps the most potentially damning would be observations of microwave anisotropies $\Delta T/T$ at levels significantly below $10^{-5}$. However, at the present time, observations of small scale anisotropy are at the level of a couple times $10^{-5}$. Observations on angular scales of degrees or more are also approaching a few $10^{-5}$.

Also not explained are the carefully reasoned conclusions of Einasto et al. (1997): ''we present evidence for a quasi-regular three dimensional network of superclusters and voids, with the regions of high density separated by 120 Mpc. If this describes the distribution of all matter (luminous and dark), then there must exist some hitherto unknown process that produces regular structure on large scales.'' Therefore the next observation that can be a potential problem for traditional models is the existence of structures with scales greater than the order of 100 Mpc, like the great wall observed by (Geller & Huchra 1989). The observations of (Broadhurst et al. 1990) show evidence for a multiplicity of such great walls with the characteristic spacing comparable to the size of the Geller-Huchra wall itself. While much debate has been made about whether or not the multiple walls of Broadhurst et al. are periodic or quasi-periodic, it does seem clear from their observations, as well as the work reported by Szalay (1990), that there is significant structure in the Universe on scales of ~100 Mpc. This is thoroughly supported by the large coherent velocity flows where the Seven Samurai and others have found evidence for the existence of an object they call the "Great Attractor" towards which the Virgo cluster and the Hydro-Centaurus cluster all seem to be flowing with a velocity ~ 600 km/sec. This again seems to indicate evidence of structures on the scales of at least 60 Mpc.

Perhaps most constraining of the traditional astronomical measurements is the existence of objects at very large redshifts. In particular, the surprising observations of Subaru telescope of the presence of galaxy clusters around $z$ ~ 6 (Ouchi et al. 2005) and a massive post-starburst galaxy at z ~ 6.5 observed by Mobasher et al. (2005). Schneider, Schmidt & Gunn (1989) have found a quasar with a redshift of 4.73 and the current record holder has $z$ = 6.28. Phinney (CalTech) was reported in Sky & Telescope to be concerned that with CDM one would not have time enough to form huge $10^9$ solar mass quasars in only one billion years, as observed. Now, from Sloan, there are more huge quasars, including one at an age of only 700 Million years. As Efstathiou and Rees (1988) have noted, if such objects are ubiquitous, this would be serious for primordial gaussian fluctuation models.

Another potentially serious observation for gaussian fluctuation models comes from the work of Bahcall & Soneira (1983), and Klypin & Khlopov (1983) where they find that clusters of galaxies seem to be more strongly correlated with each other than galaxies are correlated with each other. While Primack & Dekel (1990) have warned of the dangers of projection effects on such observations, it seems difficult to understand how projection effects would give the fractal-like behavior (Szalay & Schramm 1985). Furthermore, the southern hemisphere work of Huchra also seems to support high cluster correlations. Van den Bergh & West (1991) have also found similar correlations for the CD galaxies observed at cluster centers. The CD's should not have the projection effect problems because redshifts are known. Even Primack & Dekel now acknowledge that there seems to be some excess in cluster correlations. If such large correlations turn out to be real, they too cannot be easily explained in the gaussian model, and, as Szalay & Schramm (1985) note, they seem to be best fit by some sort of fractal-like pattern.

1.2 Late-Time cosmological Phase Transitions (LTPT)

As proposed by many researchers (Wasserman 1986; Schramm 1990; Press, Ryden & Spergel 1990; Gradwohl 1991; Frieman, Hill & Watkins 1992; Sin 1994; Sandvik, Barrow & Magueijo 2002; Nishiyama, Morita

& Morikawa 2004; etc.), Late-Time cosmological Phase Transitions (LTPTs) could be beneficial if the model assembled galaxies earlier than predicted in the cold dark matter model, and earlier formation better fit the observations. The bottom line of the theory is that if phase transitions happen after decoupling, one can avoid the constraint imposed by isotropy of microwave background.

By LTPT we will mean any non-linear growth occurring shortly after recombination. It is also possible that some normal random gaussian pattern from the very early universe could be triggered to undergo non-linear growth by some sort of phase transitions or related phenomenon occurring after recombination (Schramm 1990). In general we will see that these late-time transitions can give the smallest possible $\Delta T/T$ for a given size structure. They can produce non-gaussian structural patterns, fractal-like with large velocity flows.

As mentioned above, the very dramatic advantage of late-time transitions is that it can produce structure with $\delta\rho/\rho \geq 1$ at $z \geq 10$. Thus, one could have significant structure and a significant number of objects at high redshift, which is a problem in any normal model with the seeds forming prior to recombination. Baryonic dark matter runs into problems since it cannot get the structure we now observe formed without generating too much anisotropy in the CMB radiation. LTPTs can serve as the seeds needed to generate large scale structure after decoupling, so there is no need to the decoupling of dark matter earlier than recombination from ordinary baryonic matter.

In this paper, based on the proposed unifying theory of dark matter and quintessence I intend to reconstruct the jigsaw puzzle of structure formation in the universe. In this immense puzzle, some fragments are not discovered yet but the frame of the scenario is proposed. In the subsequent papers several aspects of the scenario are discussed. In section 2 the outline of a new scenario of large-scale structure formation was propounded. In the final section the consequences and novel predictions of the new theory are considered.

## 2  Singular-turbulent structure formation in the dark ages

2.1 Dark or bright matter?

After an inflationary epoch, the vacuum energy transforms itself to radiation energy and flows into the form of more familiar particles,

photons: 'the Creation of light'. The boson dark matter as a fluid is produced as early as this time. Surprisingly, the boson dark matter which dominates the energy density of the universe is the light itself. The background radiation energy (CMB) originates here and therefore somehow the dark matter is not dark. The boson dark matter reaches the superfluid transition temperature $T_c$, below which an electromagnetically induced ''gravity'' appears in the universe and the gravity gradually starts to work. The earliest feasible time for the transition of boson dark matter to superfluidity or BEC state is after inflation and reheating stage (Mangano et al. 2001). The temperature dependence of the Newton constant pretends to be more universal, since it does not depend on the microscopic parameters of the system (Volovik 2003). Graviton is the superfluid vacuum (condensate) of superfluid bosonic dark matter or 'light'. After the formation of quarks, dense quark matter at low temperatures is expected to be in a BCS-paired superfluid state (Iida & Baym 2002). Therefore the essence of gluon like graviton which is the superfluid vacuum of superfluid bosonic dark matter, is the superfluid vacuum of superfluid fermionic quark matter and the coherent and strongly coupled quark-gluon plasma is actually the BCS-paired superfluid quark matter. Ordinary baryonic matter acts as the 'impurities' in the superfluid boson dark matter.

As a result, after a phase transition from a symmetry state (homogeneous fluid including matter and dark matter) to a broken-symmetry state (universe with two different coexisting phases separated by an interface) at $T < T_c$, *the superfluid transition temperature of the boson dark matter*, two phases [a black brew of primordial gases ''matter'' immersed in an ocean of ''superfluid boson dark matter'', we named it ''liquid phase'', and a vacuum like phase (an immiscible phase)] and a non-zero interfacial tension appeared in the system (interfacial tension have a determinant role in the evolution of universe). The assumptions of incompressibility [Compressibility decreases the growth rate. This is an expected result, since the system has now more degrees of freedom (e.g., now the perturbation stores thermal energy as well)] and vortical turbulent velocity are justified in the liquid phase.

Therefore, in my model, after the superfluid transition of the boson dark matter and establishment of gravity in the universe, we treat the liquid phase as if it behaved like a strongly coupled, self-bound, incompressible, and non-expandable fluid [the liquid phase is treated as one single body fluid] in an expanding universe. Then, the liquid phase can no longer respond sufficiently fast to the expansion and the expansion tends to an expansion into a vacuum. We thus demand that matter in the Universe exhibits an internal attractive interaction that simulates connectivity and viscosity; that there is a tension mediating the negative hydrodynamic pressure from the Hubble expansion. Therefore the liquid phase can not exist in a state of negative pressure and in other words, hydrodynamic pressure of the liquid phase would become negative if instablity not proceeds (Bucher & Spergel 1999; Nørretranders, Bohr & Brunak 1993).

A perfect fluid with negative pressure is not possible because its sound speed would be imaginary, indicating instabilities on a short timescale whose growth rate diverges as the wavelength approaches zero. There is no elastic resistance to pure shear deformations (Bucher & Spergel 1999); therefore, the liquid phase vigorously churned by the counteraction of the gravity and universe expansion in the dark ages when $T < T_c$.

Because of the phase separation by superfluid transition of the boson dark matter, the fluctuations due to inflation will not influence mainly the process of structure formation. In our model, cosmic structure is not really a microscopic effect.

2.2 LTPT: Structure formation by Newell-Zakharov theory after decoupling and transition to superfluidity

(Silk 1973, 1974) has shown that the effects of expansion caused little deviation in turbulence from the incompressible case. We can have an incompressible turbulent medium in the liquid phase during the structure formation at $T<T_c$ and as a consequence of large Reynolds number and short hydrodynamic times, motions [on comoving scales, if compressibility increase with time] should rapidly reach the Kolmogorov spectrum.

Situation of the structure formation in the universe by the counteraction of gravity and universe expansion is alike to the excitation of surface waves (Faraday waves) and could be directly verified in a Faraday experiment. When the excitation of free surface waves exceeds a well defined threshold, the waves break and chaotic and turbulent bifurcations of Faraday surface waves lead to a low-dimensional aperiodic state with spikes, droplet ejection, and gas entrainment (Tao Shi, Goodridge & Lathrop 1997).

Parametrically forced surface waves were first studied experimentally in 1831 by Faraday. Since then, the onset of periodic surface waves and the existence of spatial and temporal chaos in this system have been extensively studied (Ciliberto & Gollub 1985; Simonelli & Gollub 1988; Meron & Procaccia 1986; Meron 1987), including the formation of quasicrystals and other wave phenomena (Tao Shi, Goodridge & Lathrop 1997). Later theoretical and experimental observations have pointed to a transition leading to cusps and singularities when the surface changes topology (Newell & Zakharov 1992). Newell & Zakharov (1992) have proposed a nonlinear singular-turbulent theory for the excitation of waves by a shear flow. This theory is based on a weak-turbulence description of the waves using kinetic equations. For energy fluxes bigger than a threshold, oscillations are excited in the instability region, which lies in the gravitational part of the spectrum. The growth of the oscillations is limited by cascade processes, as a result of which two Kolmogorov-type turbulent spectra are formed. One of them corresponds to a constant flux of wave action or of the number of waves and develops in the long-wavelength region. The other Kolmogorov spectrum corresponds to a constant energy flux directed to smaller scales:

1- For $k < k_{cr} = (g\rho_l/\gamma)^{1/2}$, where gravity dominates, the dimensionless measure of spectral energy is

$$E(k) = k_{cr}^4 \langle \eta_k^2 \rangle \sim \left(\frac{P}{P_0}\right)^{1/3} \left(\frac{k_{cr}}{k}\right)^{7/2} = E_1(k) \qquad (1)$$

$g$ is gravity, $\rho_l$ and $\gamma$ are liquid phase density and surface tension, respectively.

2- For $k > k_{cr}$, where surface tension dominates, the constant flux Kolmogorov spectrum is

$$E(k) \sim \left(\frac{P}{P_0}\right)^{1/2} \left(\frac{k_{cr}}{k}\right)^{19/4} = E_2(k) \qquad (2)$$

$\eta_k$ (Fourier transform of the liquid phase surface elevation).

$$E(k) \sim \left(\frac{k_{cr}}{k}\right)^4 = E_3(k), k < k_{cr} \qquad (3)$$

$P_0 = (\gamma g/\rho_l)^{3/4}$ is a critical value of the energy flux per unit area $P$ (for the resonant states that (Boettcher, Fineberg & Lathrop 2000) used, they found the threshold power for wave breaking to be much lower than that predicted by Newell & Zakharov. According to the Newell & Zakharov theory, for obtaining the critical power flux per unit area for breaking waves to occur a broad driving spectrum was assumed).

If during the formation of the cosmic foam, the density of the liquid phase was constant or the time scale for structure formation by Newell-Zakharov theory, $\tau_{NZ}$, is smaller than the characteristic time for the decrease in density of the liquid phase $(\rho/\dot{\rho})_f$, then the relevant equations in the liquid phase (like cascade processes as a result of two Kolmogorov-type turbulent spectra and $P_0$) remains unchanged or unaffected by universe expansion.

Because of the decelerating nature of the universe expansion, the energy flux per unit area, *P*, is a decreasing function of time.

For values $P < P_0$ and $k < k_{cr}$, the energy is carried to high, surface tension dominated wave numbers by $E_1(k)$, where the energy flux can be absorbed by surface tension wrinkling.
When $P = P_0$ will the equilibrium spectrum $E_2(k)$ be exactly realized because only then is $E_1(k_{cr}) = E_2(k_{cr})$.
For $P < P_0$ and $k > k_{cr}$, $E(k) > E_2(k)$.

For this range of fluxes the interface of two phases stays smooth, energy is transferred to scales where viscosity is important by wave-wave interactions and the topological boundary condition that a liquid particle on the surface stays there remains intact.

$$\text{For } P > P_0, k_l < k_{\text{cr}}, \quad k = k_l = k_{cr}\left(\frac{P_0}{P}\right)^{2/3} < k_{cr} \qquad (4)$$

It is interesting that the criterion that $P > P_0$ is also the criterion for the Kolmogorov–Zakharov spectrum to meet the Phillip's spectrum before $k = k_{\text{cr}}$. $P$ is too much for a smooth surface to handle. In order to absorb the energy flux, the surface must increase its area. But it cannot use a smooth surface on which surface tension waves redistribute the energy to smaller scales because $k_l < k_{cr}$. The only remaining way for the surface to achieve a greater area is for it to break and to spray droplets of liquid (matter & dark matter) into the phase immediately above the interface causing the formation of cosmic foam consisting of liquid droplets of a size at which surface tension effects are important. Because of that $P_0$ is a constant value or an increasing function of time and $P$ is a decreasing function of time, so if $P > P_0$ early, soon or late, $P$ becomes smaller than $P_0$ and the cosmic foam formation is stopped. If $P$ is strong enough ($P > P_0$) early in the formation of cosmic foam (according to our theory, the liquid phase vigorously churned in the dark ages), then because of the short $\tau_{\text{NZ}}$ (because of the collapsing singular nature of the process which leading to the breakdown of the interface) and long characteristic time (in comparison with $\tau_{\text{NZ}}$) for the decrease in $P$ ($P/\dot{P}$), before that $P$ becomes smaller than $P_0$, a fully developed, spatially homogeneous cosmic foam can form, so we assume the cosmic foam to be distributed uniformly in the interface of two phases. In the course of the expansion of the Universe and via a top-down scenario, the large-scale structure is formed by the present epoch. Earlier interpretations of the large scale structure as Voronoi foam are primarily based on the geometrical similarity; our model is rooted in a physical mechanism applicable to the cosmological problem and resulting morphology is consistent with the observed correlation functions, since a Voronoi tessellation in its original sense has happened in our model. The initial spherical bubbles (voids) assembled and collided on the interface of two phases and because of the dynamical evolution of the foam a cosmic foam where the voids retain only approximate spherical formed as convex polyhedrons.

In the Newell-Zakharov theory, in the gravitational-capillary transition region the spectrum with a constant energy flux must be joined to the Kolmogorov spectrum of the capillary waves.

Estimation of cosmic foam thickness and liquid/other phase droplet/bubble size
Assumptions:
1. The various forms of energy, surface, potential, and kinetic, have the same orders of magnitude.
2. All available energy goes into various forms of energy in the foam, roughly on an equal basis.

Equating the surface and potential energy, that is, assuming that the potential energy of a liquid droplet is balanced by its surface energy, we obtain, $\gamma\lambda^2/\rho_1 \sim g\lambda^3 h$.

For small times, we equate the surface energy of the foam in a column of height $h$ and unit cross section to the energy input, namely, $(\gamma\lambda^2/\rho_1)(h/\lambda^3) \sim Pt$, which leads to the laws

$$h \sim (Pt/g)^{1/2}, \lambda \sim \gamma/\rho_1(gPt)^{1/2} \tag{5}$$

$\lambda h = k_{cr}^{-2} \rightarrow$ the geometric mean of foam thickness and liquid droplet size is $k_{cr}^{-1}$, the scale at which gravity and surface tension effects balance. $\lambda$ is the size of the droplets/bubbles where all the energy flux can be absorbed which affects the sequential fragmentation of voids (bubbles). $h$ is a decreasing function of time and $\lambda$ is an increasing function of time although during the small time of structure formation, the difference is negligible.

In the absence of viscosity, the only relevant time scale is $(\rho_1\lambda^3/\gamma)^{1/2}$ and replacing the $t$ in equation (5) by this value, we obtain

$$hk_{cr} \sim \left(\frac{P}{P_0}\right)^{2/7} > 1, \lambda k_{cr} \sim \left(\frac{P}{P_0}\right)^{-2/7} < 1 \tag{6}$$

Note that $h$ is larger than the critical wavelength $k_{cr}^{-1}$ but smaller than the scale $k_1^{-1}$ of waves which produce the spray,

$$k = k_1 = k_{cr}\left(\frac{P_0}{P}\right)^{2/3} < k_{cr} \tag{7}$$

$$hk_1 \sim \left(\frac{P}{P_0}\right)^{-8/21} < 1 \qquad (8)$$

Wave collapse or breaking is the most effective mechanism of the wave-energy dissipation. For the liquid phase surface waves the analogous phenomenon leads to the infinite second derivative of the surface profile (so that angles or cones appear on the surface). Checking analyticity violation is the most sensitive tool for studying that set of collapses. Loss of analyticity of vortex sheets at the nonlinear stage of the Kelvin-Helmholtz instability is such an example. It was assumed that the singularity formation on the free surface of the ideal fluid or in a more general case, for the boundary between two ideal fluids, is mainly connected with inertial forces; other factors give minor correction (Kuznetsov, Spector & Zakharov 1994). Wave-breaking states necessarily contain singularities, since they exhibit a change from a simply connected free surface to a free surface connected in multiple ways with droplet ejection and entrainment of the other phase. Waves on the open fluid that break can also show several types of singular phenomena. The spray and foam produced in liquid phase breaking also contain gravity-capillary singular phenomena associated with the change in topology. This type of spray production also occurs in rivers and streams with significant surface and bulk turbulence. Therefore during the wave-breaking process, in addition to the explosion of liquid phase (including matter and dark-matter) into droplets, one can also have entrainment of the other phase into the liquid phase and the formation of a cloud of other phase bubbles (voids) surrounded by a connected fluid region. For this situation, and for a combination of other immiscible phase bubbles (voids) in liquid phase and liquid phase bubbles (droplets) in immiscible phase, similar calculations would obtain.

In addition to the multiphase possibilities, our model must take account of the turbulence in the liquid phase itself.

A harmonic acceleration, possibly in the presence of a constant body force, associated with Faraday waves (Wright, Yon & Pozrikidis 2000). Droplet ejecting Faraday waves (surface wave singularities) are produced by vertically oscillating a fluid surface with sufficient acceleration. The flat surface becomes unstable to periodic surface waves at a critical acceleration (via the Faraday instability). As the excitation is increased further, we observe a sharp transition to a state with spikes on the surface which eject droplets from the tip (Goodridge, Tao Shi & Lathrop 1996;

Tao Shi, Goodridge & Lathrop 1997; Saylor & Handler 1997; Hogrefe et al. 1998; Goodridge, Hentschel & Lathrop 1999).

As mentioned above, the ejection is preceded, by one period; by a rounded wave peak whose amplitude increases until the slope becomes infinite and a slight overhang is formed. This last smooth wave collapses into a sharp-cornered depression, which then focuses to form a growing spike. The spike increases in amplitude and width and suffers droplet producing Rayleigh instabilities (which cause the wave tips to break under the influence of surface tension forces) near the cylindrical top. Finally, it collapses due to gravitational forces, leaving behind a stretched neck which also breaks into droplets. The collapsing spike often entrains bubbles into the fluid bulk (Goodridge, Tao Shi & Lathrop 1996). In low-viscosity liquids, spikes are produced which immediately break up into droplets. In high-viscosity fluids, these peaks maintain their structure and can produce long filaments before droplet breakoff occurs (Goodridge et al. 1997). The rate of breaking events approaches zero gradually with decreasing acceleration. (Goodridge, Hentschel & Lathrop 1999) experimentally support the hypothesis put forward by (Newell and Zakharov 1992) that a continuous transition exists from unbroken surfaces to surfaces with droplets and spray.

Well-controlled experiments exhibiting droplet ejection are Faraday waves forced above a threshold acceleration (Goodridge, Tao Shi & Lathrop 1996). The flux $P$ can be easily controlled and the energy can be injected at whatever wave number $k < k_0$ found to be suitable. Moreover, the surface tension can also be sensitively controlled so that a fully developed foam is formed (Newell & Zakharov 1992). Faraday waves have been well studied and the ejection threshold has been characterized over a wide parametric range (Goodridge et al. 1997). Droplet ejection in Faraday waves is a random and uncorrelated phenomenon. Individual waves eject independently of other waves in a fashion similar to a radioactive decay (Goodridge, Hentschel & Lathrop 1999).

Droplet ejection occurs in waves restored by gravitational forces (lower frequency gravity waves) and those restored by surface tension forces (higher frequency capillary waves). It is notable that although the excitation occurs in gravity length scales the fluid motion transfers energy to capillary length scales in the spike (Goodridge, Tao Shi & Lathrop 1996).

A theory, motivated by the cylindrical shape of the pre-singularity surface depression in the Faraday experiment, predicted that the resulting shape of the singularity would grow according to the power law $z = btr^{-}$

$^{1/2}$. Investigation of the Faraday experiment resulted in a confirmation of the theory. Furthermore, information was derived about the velocity field of the fluid after the singularity occurred by comparing the experimental value for *b* to its predicted form (Hogrefe et al. 1998).

Droplet ejection from liquid surfaces is a ubiquitous phenomenon in nature. Wave breaking on the surfaces of oceans and lakes, the spray from turbulent rivers, and the splash from a raindrop hitting a puddle all involve the production of small, energetic droplets which escape and then rejoin the main body of the liquid. These surface waves receive the energy needed to create droplets from sources such as the shear flow or gravity driven flow and require a certain minimum energy flux to produce droplets (Goodridge et al. 1997). Viscous dissipation may be the dominant sink of energy for very short waves, but breaking affects all part of the spectrum. It is probably the main cause of long gravity wave attenuation. Waves break when inertial accelerations exceed the restoring force, or when particle velocities at the wave crest outstrip the phase velocity of the wave. Wave breaking manifests itself through whitecaps. Breaking serves to limiting height of surface waves, dissipating surface-wave energy, some of which is available for turbulent mixing. Breaking is multi-scale process: from large-scale breaking waves and smaller-scale waves breaking on longer waves to micro-scale breaking (The wave breaking process is pictured as highly nonlinear in wave steepness, having no effect until some limiting steepness is reached when the wave becomes unstable and spills or plunges forward, producing whitecaps at large scales or a micro breaker at small scales. At the end of the breaking event a substantial energy loss may occur). Direct measurements show that breaking generates spectra of the intensive outbursts of turbulence with dissipation rate that is orders of magnitude larger than the mean value. These events result in the roughly lognormal probability distribution of the dissipation rate.

Surface tension as restoring force

From (Kolmogorov 1949), we have:
$$\upsilon_d \approx \left(\frac{\nu\, d}{\lambda_0^2}\right) \text{ or } \upsilon_d \approx \left(\frac{\nu\, d^{1/3}}{\lambda_0^{4/3}}\right) \tag{9}$$

Owing to the surface tension, the jet of the immiscible phase disintegrates into bubbles, the bubbles are broken to a certain limit, and the bubbles of sufficiently small diameter *d* are preserved, since for small *d* the breaking forces acting on them due to the velocity differences, which are of the order of $\upsilon_d$, are small for small *d* and can no longer overcome the surface tension; or When the turbulent stresses are equal to the confining stresses, $\tau_t(d) = \tau_s(d)$, a critical diameter, $d_c$, is defined such that particles with $d < d_c$ are stable and will never break (Kolmogorov 1949; Hinze 1955). A particle of size $d > d_c$ has a surface energy smaller than the deformation energy ($\tau_t(d) > \tau_s(d)$), and thus, the particle deforms and eventually breaks up.

Table 1. Summary of dimensionless characteristics (Kolmogorov 1949)

|  | $v' \ll v$ or $v' \approx v$ | $v' \gg v$ |
|---|---|---|
| $d \ll \lambda_0$ | $We_d$ and $v'/v$ | |
| $d \approx \lambda_0$ | $We_d$, $d/\lambda_0$ and $v'/v$ | |
| $d \gg \lambda_0$ | $We_d$ | $We_d$ and $Re'_d$ |
| $d \gg \lambda'_0$ | - | $We_d$ |

$We_d = \sigma/\upsilon_d^2 d\rho$ (Weber number)
$\lambda_0 = (v^3/\varepsilon)^{1/4}$ (Kolmogorov dissipation length or Internal scale)
For a tube of diameter $D$, outside the laminar boundary layer:
$\lambda_0 = g(r/D) (v^3 D/u_*^3)^{1/4}$
$\lambda'_0 = (v'/v)^{3/4} \lambda_0$.

Under conditions of very rapid shear, instabilities of the Rayleigh type may be important in foam formation by beating or shaking.

Origin of lognormal droplet/void size distribution

In the past decade, if many papers deal with the 3D distribution of galaxies very few explore how the voids are distributed. Only recently, the void size distribution as an indicator of the dynamics of void formation is considered. If the resemblance of large scale structure to a cosmic foam be further than a geometrical analogy, then as in hydrodynamic foams, the distribution of void size is a function of the method of foam production (Weaire & Hutzler 1999). With a great degree of accuracy, the PDF of void size in different large surveys like LCRS (Müller et al. 2000) and 2dFGRS (Figure 1) is lognormal. This distribution is considered for the first time by Zaninetti (1991) in order to simulate the CFA data. As proposed by (Kolmogorov 1941, 1949), a foam with a lognormal PDF is produced in a turbulent medium in which bubbles (voids) are repeatedly fragmented. Therefore the observed size distribution of voids in redshift surveys as in ancient cosmogonies further support the turbulent nature of structure formation in the universe! We can interpret the observed log-normal void size distribution as a consequence of a mixing process, in which turbulent mixing at a material interface between two phases is essential to the evolution of the large-scale structure.

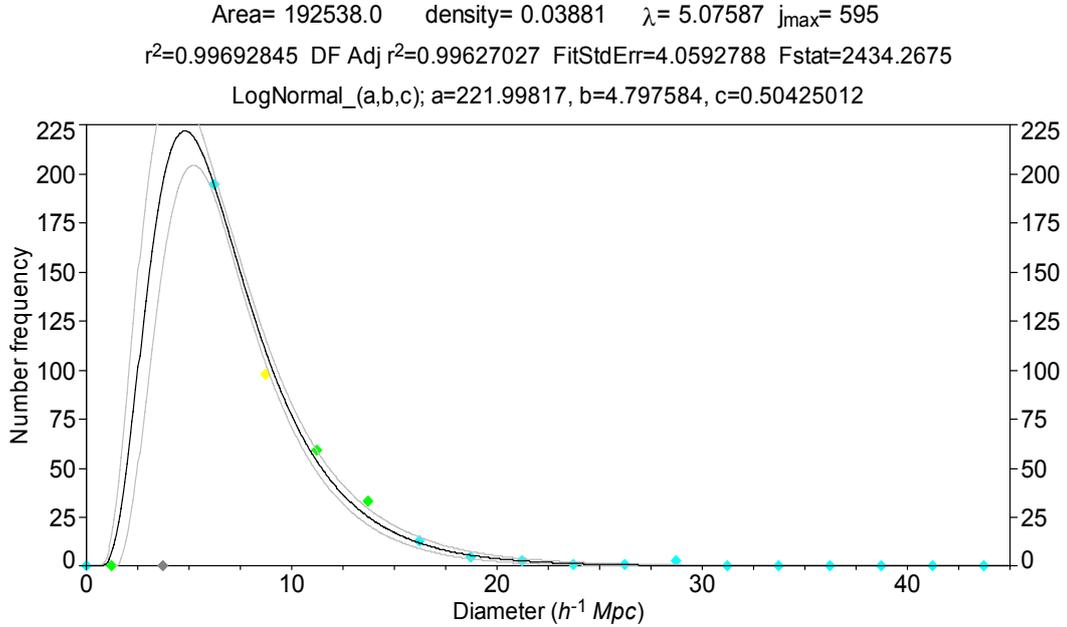

Figure 1. Lognormal void size distribution (2dFGRS volume limited subset). The mean galaxy separation ($\lambda = 5.07587$ $h^{-1}$ Mpc) corresponds to the lower limit on void size (truncation size).

According to the (Brovchenko & Maderich 2004), almost all statistical models of break-up of an immiscible fluid immersed into a turbulent flow were not able to reproduce observed distribution of oil droplet size entrained by breaking waves in stormy conditions. Instead, the new model of the breakup based on (Kolmogorov 1941) approach was proposed to reproduce observed lognormal distribution of oil droplet sizes. The theory of break-up of bubbles immersed into a turbulent flow in principle is no different from that of drops. The only actual difference is the fact that the critical size of the bubble at which break-up occurs differs from that of drops, so we can use (Brovchenko & Maderich 2004)'s results in the case of the breakup of entrained immiscible phase in the surface layer of the liquid phase.

For a given *P*, the droplets/bubbles will continue to break up by surface deformation and form even smaller droplets/bubbles until they reach a size λ where all the energy flux can be absorbed. At this stage, the cosmic foam will cease to grow and the liquid droplets will simply oscillate. If viscosity is present, we can form the Kolmogorov length scale (internal scale)

$$l = \frac{\nu}{P^{1/3}} = \frac{\nu}{P_0^{1/3}} \left(\frac{P_0}{P}\right)^{1/3} \qquad (10)$$

so *l* is an increasing function of time.
However, the ratio

$$\frac{l}{\lambda} = \frac{\nu \rho_1^{3/4} g^{1/4}}{\gamma^{3/4}} \left(\frac{P_0}{P}\right)^{1/21} \qquad (11)$$

becomes smaller as *P* increases, so that even though the bubble size decreases, the Kolmogorov length scale, at which viscous effects would be expected to become important, decreases at a slightly faster rate. Thus the mechanism for dissipation of the cosmic foam must involve more complicated dynamical processes such as a weak turbulence energy transfer by small amplitude waves on the bubble surfaces.

Late Time Phase Transition

In the case of cosmic foam formation, late time phase transition causes a change of topology. Order parameter - in the model - is the deviation of the liquid phase surface (including matter and dark-matter) from the smoothness and connectivity it has at $P < P_0$, the critical value of the energy flux per unit area. At $P > P_0$ the smooth surface condition is broken and cusps and singularities (droplets, spray, and foam, produced by breaking deep fluid waves are examples of dynamical singularities on the free surface of a fluid) are formed. The symmetry breaking is spontaneous. A phase transition from a symmetry state (unbroken surface or turbulent nonejecting state) to a broken-symmetry state (surface with droplets and spray or turbulent ejecting state) was causing the formation of an [liquid phase (matter & dark matter) — other immiscible phase] foam.
Under hurricane-like conditions in the beginning of the dark ages which lead to uniform distribution of cosmic foam, with considering the effect of long waves (as a "geometric imperfection") which makes the phase transition continuous (Newell & Zakharov 1992), we suggest that the behavior near $P = P_0$ is equivalent to a second order phase transition (Newell & Zakharov 1992; Goodridge, Hentschel & Lathrop 1999); thus geometric imperfections have a determinant role in resolving the order of phase transition. The phase transition is influenced by several factors: (1) the kinematic surface tension $\gamma/\rho$, (2) the kinematic viscosity of the fluid $\nu = \mu/\rho$, and (3) the applied forcing frequency $\omega_0$. Low-viscosity fluids have threshold accelerations which depend on only surface tension and forcing frequency.

Turbulence in superfluids

Turbulence in superfluids is governed by two dimensionless parameters. One of them is the intrinsic parameter $q = \alpha/(1 - \alpha')$ (dimensionless parameters $\alpha'$ and $\alpha$ come from the reactive and dissipative forces acting on a vortex when it moves with respect to the normal component) which characterizes the friction forces acting on a vortex moving with respect to the normal component, with $q^{-1}$ playing the same role as the Reynolds number $Re = UR/\nu$ in classical hydrodynamics. The developed turbulence described by Kolmogorov cascade occurs when $Re \gg 1$ in classical hydrodynamics, and $q \ll 1$ in the superfluid hydrodynamics. Another parameter of the superfluid turbulence is the superfluid Reynolds number $Re_s = UR/\kappa$, which contains the circulation quantum $\kappa$ characterizing quantized vorticity in superfluids. This parameter may regulate the crossover or transition between two classes of superfluid turbulence: (i) the classical regime of Kolmogorov cascade where vortices are locally polarized and the quantization of vorticity is not important; and (ii) the quantum Vinen turbulence whose properties are determined by the quantization of vorticity. The phase diagram of the dynamical vortex states is suggested (Volovik 2003).

The turbulence in the superfluid component with the normal component at rest is referred to as the superfluid turbulence (Volovik 2003). The important feature of the superfluid turbulence is that the vorticity of the superfluid component is quantized in terms of the elementary circulation quantum $\kappa$. So the superfluid turbulence is the chaotic motion of well determined and well separated vortex filaments. The further simplification comes from the fact that the dissipation of the vortex motion is not due to the viscosity term in the Navier-Stokes equation which is proportional to the velocity gradients $\nabla^2 \mathbf{v}$ in classical liquid, but due to the friction force acting on the vortex when it moves with respect to the normal component. This force is proportional to velocity of the vortex, and thus the complications resulting from the $\nabla^2 \mathbf{v}$ term are avoided.

$$v_r = (\varepsilon r)^{1/3} \tag{12}$$

This must be valid both in classical and superfluid liquids. What is different is the parameter $\varepsilon$: it is determined by the dissipation mechanism which is different in two liquids.

Instead of $\varepsilon = \nu v_{r0}/r_0^2$ in classical liquids, we have now

$$\varepsilon \sim q\,U^2\,\varepsilon^{1/3} r_0^{-2/3},\ r_0 \sim q^{3/2} R,\ v_{r0} \sim q^{1/2} U \tag{13}$$

As in the Kolmogorov cascade for the classical liquid, in the Kolmogorov cascade of superfluid turbulence the dissipation is concentrated at small scales,

$$\varepsilon \sim qU^2\, v_{r0}/r_0 \tag{14}$$

while the kinetic energy is concentrated at large scale of container size:

$$E = (\varepsilon R)^{2/3} = U^2 \tag{15}$$

The dispersion of the turbulent energy in the momentum space is the same as in classical liquid

$$E = \int_{r_0}^{R} \frac{dr}{r}(\varepsilon r)^{2/3} = \int_{k_0}^{1/R} \frac{dk}{k}\frac{\varepsilon^{2/3}}{k^{2/3}} = \int_{k_0}^{1/R} dk E(k),$$

$$E(k) = \varepsilon^{2/3} k^{-5/3} \tag{16}$$

As distinct from the classical liquid where $k_0$ is determined by viscosity, in the superfluid turbulence the cut-off $k_0$ is determined by mutual friction parameter $q$: $k_0 = 1/r_0 = R^{-1} q^{-3/2}$.

At a very small $q$ the quantization of circulation becomes important. The condition of the above consideration is that the relevant circulation can be considered as continuous, i.e. the circulation at the scale $r_0$ is larger than the circulation quantum: $v_{r0} r_0 > \kappa$. This gives

$$v_{r0} r_0 = q^2 UR = q^2 \kappa \mathrm{Re}_s > \kappa,\ \mathrm{Re}_s = UR/\kappa \tag{17}$$

i.e. the constraint for the application of the Kolmogorov cascade is

$$\mathrm{Re}_s > 1/q^2 \gg 1 \tag{18}$$

So, the crossover between the classical and quantum regimes of the turbulent states occurs at $\mathrm{Re}_s q^2 = 1$ and the Vinen state which probably occurs when $\mathrm{Re}_s q^2 < 1$.

The transition (or maybe crossover) is suggested here between the quantum and classical regimes of the developed superfluid turbulence, though there are arguments that the classical regime can never be reached because the vortex stretching is missing in the superfluid turbulence (Kivotides & Leonard 2003); although, (Kivotides *et al.* 2002) study numerically statistics of superfluid turbulence. They generate a quantized superfluid vortex tangle driven by a realistic model of normal-fluid turbulence whose energy spectrum obeys Kolmogorov classical $k^{-5/3}$ law, where $k$ is the wave number. They find that the resulting superfluid velocity spectrum has approximately a $k^{-1}$-dependence for wave numbers of the order of $1/\delta$ and larger, where $\delta$ is the average intervortex spacing. This result is similar to what happens in a pure superflow (Araki, Tsubota

& Nemirovskii 2002). (Kivotides *et al.* 2002) also find that the spectrum of the total velocity field follows the classical $k^{-5/3}$ law, even at temperatures low enough that the normal-fluid mass is only 5% of the total helium mass.

Tsubota & Kobayashi (2005) introduce an energy injection at large scales as well as the small-scale dissipation, and obtain the statistically steady turbulence made by the balance of the injection and the dissipation. The inertial range still takes the Kolmogorov spectrum for the incompressible kinetic energy. The energy flux, which transfers the energy from large to small scales, is almost constant in time and independent of the wave number, being consistent with the energy dissipation rate at small scale. These discoveries show surprising properties of the inertial range of QT which have ever been unclear. The inertial range of QT is also sustained by the Richardson cascade process of quantized vortices. Kobayashi & Tsubota (2005), in their experiments, show a similarity between ST and CT. This can be understood using the idea that the superfluid and the normal fluid are likely to be coupled together by the mutual friction between them and thus to behave like a conventional fluid. Since the normal fluid is negligible at very low temperatures, an important question arises: even without the normal fluid, is ST still similar to CT or not? Quantum turbulence (QT) consisting of quantized vortices can propose a prototype of turbulence much simpler than classical turbulence. One of the points is how QT can reproduce the essence of classical turbulence. Recently they made the numerical analysis of the Gross-Pitaevskii equation with a dissipation term that works only in the scale smaller than the healing length, thus succeeding in obtaining clearly the Kolmogorov spectrum which is one of the most important statistical laws in turbulence.

The turbulence in classical liquids is thought to be characterized by the dynamics of the vortex tubes, whose radii are of order of the dissipative Kolmogorov scale. In some regime, the superfluid turbulence is similar to that in classical liquids with modified dissipation. Thus the quantum liquid serves as a physically motivated example of the liquid with the non-canonical dissipation, which requires the general analysis of different models of dissipation and forcing.

As proposed by (Seidel & Maris 1994) the formation of structure by NZ theory on the surface of a levitated superfluid droplet, is possible. Because the surface of a levitated droplet has no boundaries, it is ideally-suited for the study of non-linear interactions of capillary waves based on NZ theory.

Newell-Zakharov theory is based on a weak-turbulence description of the waves using kinetic equations. The growth of the oscillations is limited by cascade processes, as a result of which two Kolmogorov-type turbulent spectra are formed. So, if according to (Araki, Tsubota & Nemirovskii 2002; Kobayashi & Tsubota 2005; Kivotides *et al.* 2002), the classical regime of Kolmogorov cascade can be reached in the superfluid dark matter ($\rho_n \ll \rho_s$) then there is no constraint on the ratio $\rho_s/\rho$ for the cosmic foam formation based on NZ; otherwise there is a constraint on the superfluid part of dark matter because for the structure formation based on NZ and also sequential fragmentation of voids in a turbulent medium, the development of the Kolmogorov turbulent energy cascade is essential.

Cosmic foam is produced by agitation and the large-scale structure was born on the surface of a churning liquid phase. The voids (bubbles) produced under conditions where strong instabilities in the interface of two phases are the main mechanism for bubble (void) production. Such instabilities are believed to occur during liquid wave break-ups or agitation of liquid-gas mixtures. Wave break-up fed by shaking energy is probably the main mechanism responsible to foam creation.

2.3 Turbulent galaxy formation and clustering

(Gamov 1954) suggested that the space and mass distributions of galaxies were fossils of powerful primordial turbulence driven by the Big Bang because density fluctuations of the turbulence would influence the formation of such gravitational structures. (Brown 1985) explored the idea that galaxies might have resulted from eddies in a turbulent early universe. He has explored the galaxy-forming potential of an early universe that fragmented in the presence of large scale shearing flows arising from turbulence. He also concludes that the dark matter must have accompanied the luminous matter in the viscous evolution in order to maintain the required gravitational potential.

Fragmentation is a ubiquitous phenomenon in the structure formation of the Universe from the sequential fragmentation of the voids to the formation of the first galaxies and stars.

On the basis of our top-down model, after the formation of cosmic foam layer, a probable and plausible scenario for turbulent galaxy formation is:
Dark matter flow or coherent motions can be induced by the fluid dynamics of the cosmic foam itself (such as cosmic foam drainage, local fluid flow, etc.). Therefore in this case the coherent motions in foam sheets become turbulent and the protogalaxies and the first massive galaxies will form from coherent structures like eddies, vortices, etc.: *Coherent structures emerge from chaos, under the action of an external constraint (instability of the inflectional basic velocity profile).*

Turbulence always starts at small scales and cascades to large, therefore the formation of the clusters of galaxies in foam sheets is bottom-up. (Krishan & Sivaram 1991) and (Prabhu & Krishan 1994) showed that the clustering and superclustering of galaxies and clusters respectively could be viewed as the outcome of the *''inverse cascade''* process in an incompressible turbulent medium. Thus, the universe is a hierarchy of eddies. Eddies towards the small scale end can be identified with galaxies and those towards the large, the superclusters (Figs. 2 and 3). Here we suggest a physical mechanism based on inverse cascading which quite naturally yields a bottom-up hierarchical structure.

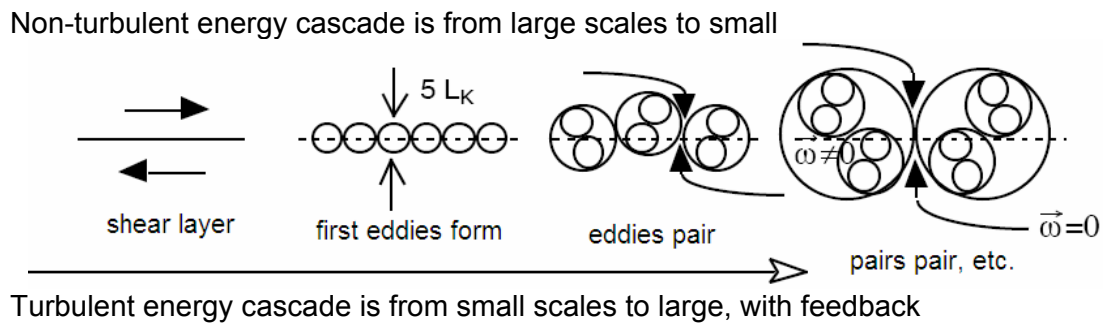

Figure 2. Schematic of the turbulence cascade process, from small scales to large (Gibson 1999).

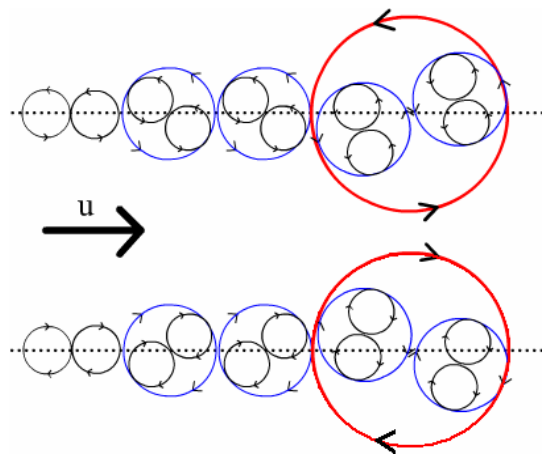

Figure 3. Hierarchy of eddies and inverse cascade process.

In order for any physical process to operate in an expanding universe, its characteristic time scale must be shorter than $H^{-1}(t)$. In the case of turbulence, the relevant time scale is that of the nonlinear interaction among eddies $l/v(l)$, where $l = a(t)\pi/k$ is the physical size corresponding to the wavenumber $k$ ($k$ is a wavenumber defined with respect to the *comoving* coordinates $x^i$), $a(t)$ is cosmic scale factor or the expansion parameter, and $v(l)$ is given by $\frac{1}{2}v^2(k) = \int_k^\infty E(k')dk'$, where $v(k)$ is the turbulent velocity corresponding to $k$ and $E(k)$ is the turbulent kinetic energy spectrum; so that the above condition becomes
$l/v(l) \leq H^{-1}$

In transition to superfluidity, a dense system of strongly interacting particles can be represented in the low-energy corner by a dilute system of weakly interacting "elementary excitations" or "quasiparticles" (Khalatnikov 1965). During and after the formation of turbulent coherent structures the compressibility was increased. According to (Wetterich 2003) a bound $|R| = \left|\frac{\Delta\alpha(z=0.13)}{\Delta\alpha(z=2)}\right| < 0.02$ strongly favors quintessence with a time varying equation of state $w = p/\rho$, where the value of $(1+w)$ at present is substantially smaller than for $z = 2$ and equivalently the velocity of sound in the medium is reduced (compressibility is increased) and this process has the main influence on the process of structure formation (transition to gravitational instability). It is possible that this monotonically increasing trend of $w$ extend to high-redshifts. Then nonlinear gravitational, cooling and thin-shell instabilities may play significant roles in the fragmentation of cosmological sheets (Annios, Norman & Anninos 1995). So the sheets fragmented under hydrodynamical and gravitational instabilities and the first stars form by the interplay between supersonic turbulence and self-gravity (Klessen 2001). When the energy density of the BEC or superfluid part of DM exceeds some critical value, the condensates rapidly collapses on the turbulent coherent structures (maybe into some structures like compact boson stars and black holes) and forms SFDM (equivalent of ΛCDM) which work as the standard cold dark matter halo of galaxies. Dark matter coherence manifests itself in a rich variety of phenomena, namely, electromagnetically induced transparency (EIT) or ''dark resonance''; therefore we can conclude that dark matter was bright before its transition to coherent state and after that shadowed itself in a veil and it will enlighten the Universe again.

## 3 Consequences of the model

Results and consequences of the model are listed below and some newly claimed outcomes are described:

- Comprehensive model of structure formation (compatibility with ΛCDM with all the successes of that model without discrepancies on galactic scales)
- Top-down singular & turbulent structure formation which results in a bottom-up hierarchical clustering
- New insights into the essences of: wave-particle duality of matter, light, gravity and inertia, SCQGP, CMB, and boson dark matter as the light itself
- Time varying quintessence and dark matter
- Unified model for quintessence and dark matter
- Thermodynamic instability of cosmic foam
- Strong intra-cluster motions & the fluid dynamics of the cosmic foam (large coherent velocity flows)
- $\frac{\delta T}{T} \leq 10^{-5}$
- Structures $\geq 100 Mpc$
- Objects existing at $z \geq 5$
- Strong non-linearity is formed in the very early stage, therefore no extra biasing process is needed
- Large cluster-cluster correlations
- Precise determination of the observed void size distribution
- Ripple-like fine structures in slightly increasing rotation curves of galaxies
- Central object of galaxies was formed at the same time than the halo which better fit the new observations
- Presence of viscosity in the structural development of universe
- Prevention from shock wave problem after decoupling
- Topology of the Universe
- Aether and Newton philosophy (motion of the earth through the aether)
- Expansion of the Earth, essence of mounts

3.1 Strong intra-cluster motions and the fluid dynamics of the cosmic foam (large coherent velocity flows)

The peculiar velocities originate from motions of galaxies within gravitationally bound systems but mainly from strong coherent motions in matter because of dynamical evolution of the cosmic foam such as cosmic foam drainage, local fluid flow, flow through Plateau border junctions (in liquid foams, the liquid largely resides in a network of vertices and edges –the Plateau borders. Maybe the cosmological attractors are Plateau borders); (Figure 4). For example, the Great Attractor seems to be going with the flow rather than causing it (Mathewson & Ford 1994). The flow is not uniform over the Great Attractor region. It seems to be associated with the denser regions which participate in the flow of amplitude about 400 km $s^{-1}$. In the less dense regions, the flow is small or nonexistent. This makes the flow quite asymmetric and inconsistent with that expected from large-scale, parallel streaming flow that includes all galaxies out to 6000 km $s^{-1}$ as previously thought. The flow cannot be modeled by a Great Attractor at 4300 km $s^{-1}$ or the Centaurus clusters at 3500 km $s^{-1}$. Indeed, from the density maps derived from the redshift surveys of "optical" and *IRAS* galaxies, it is difficult to see how the mass concentrations can be responsible particularly as they themselves participate in the flow. These results bring into question the generally accepted reason for the peculiar velocities of galaxies that they arise solely as a consequence of infall into the dense regions of the universe. Also, Hudson (1994) finds that the Centaurus-Hydra-Virgo and Pave supercluster complex is not primarily responsible for the large streaming motions of galaxies. Hudson believes that most of the bulk motion of the 405 km $s^{-1}$ which is required to agree with the predicted motion of the Local Group is due to sources beyond 8000 km $s^{-1}$. The main difficulty that the standard theory faces is to explain why the large visible mass centers of the Local Universe do not appear to produce large-scale flows but instead fully participate in the flows themselves. Has the flow of the Great Wall been detected?

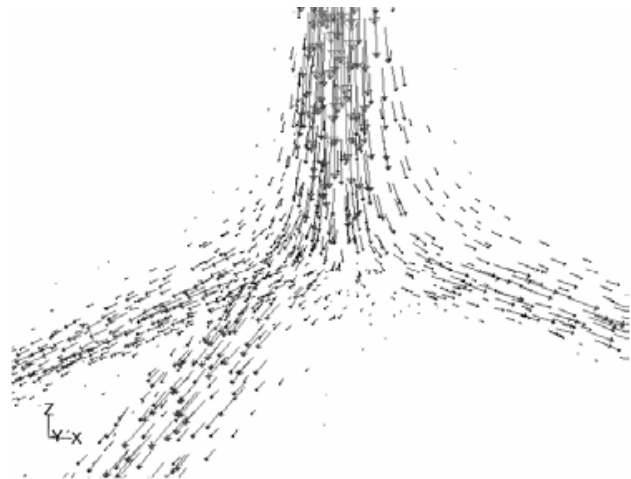

Figure 4. Fluid dynamics and coherent motions of the cosmic foam

## 3.2 Thermodynamic instability of cosmic foam

Many naturally occurring structures, from the common soap froth to the large-scale distribution of galaxies in the universe, consist of statistically homogeneous domains separated from each other by distinct boundaries. These boundaries are associated with an interfacial energy (surface energy in three dimensions, wall energy in two dimensions). If the total energy is simply the product of the boundary area times a 'surface tension' or surface energy, any reduction in total interfacial area will reduce the energy. Hence such structures are intrinsically unstable, always evolving towards patterns with less surface area, unless other factors (such as boundary pinning or short-range repulsive forces) intervene. The basic mechanism to reduce interfacial area is the elimination of entire domains (Glazier & Weaire 1992).

## 3.3 Precise determination of observed void size distribution (lognormal void size distribution)

One of the important differences between other models of structure formation and my model (based on the hydrodynamic instability) is the predicted void size distribution; our model, based on the break up of the voids in a turbulent regime, predicts the observed lognormal void size distribution, but the simulated PDFs of void size in other models (expansion of underdensities or explosion models) have some degree of disagreement to the observations.

## 3.4 Prevention from shock wave problem of plasma turbulence theories

Our model helps to avoid one serious problem of all plasma turbulence models including that of Goldman & Canuto (1993): For turbulent velocities, in other models, one must expect the generation of shock waves at large scales because of the decrease in the $c_{sound}$ at decoupling epoch. Since these are not observed (Peebles 1980, 1993). In our model based on LTPTs, the problem will be avoided.

## ACKNOWLEDGMENT
I thank Sepehr Arbabi Bidgoli for wide-ranging discussions.